\documentclass[titlepage, twocolumn, 12pt]{report}

\usepackage[newfloat,frozencache,cachedir=.]{minted}
\usepackage[utf8]{inputenc}
\usepackage{blindtext}
\usepackage{capt-of}
\usepackage[top=1in, bottom=1in, left=1in, right=1in]{geometry}
\usepackage{graphicx}
\usepackage{lipsum}
\usepackage{makecell}
\usepackage{multirow}
\usepackage{url}

\graphicspath{ {./images/} }

\newenvironment{Figure}
  {\par\medskip\noindent\minipage{\linewidth}}
  {\endminipage\par\medskip}

\newcommand{\fig}[3]{
    \begin{Figure}
        \centering
        \includegraphics[width=\linewidth]{#1}
        \textbf{\captionof{figure}{#2}}
        \label{fig:#3}
    \end{Figure}
}

\title{A Brief Analysis of the\\Apollo Guidance Computer}
\author{
    Charles Averill \\
    \texttt{charles.averill@utdallas.edu}
}
\date{Fall 2021}

\begin{document}

\maketitle

\clearpage

\section{Overview}

The \textbf{AGC}\footnote{This report will cover the Block II iteration of the AGC. The first iteration, Block I, was not used in lunar flight.} was designed with the sole purpose of providing navigational guidance and spacecraft control during the Apollo program throughout the 1960s and early 1970s. The AGC sported 72kb of \textbf{ROM}, 4kb of \textbf{RAM}, and a whopping 14,245 \textbf{FLOPS}, roughly 30 million times fewer than the computer this report is being written on. 

These limitations are what make the AGC so interesting, as its programmers had to ration each individual word of memory due to the bulk of memory technology of the time. Despite these limitations (or perhaps due to them), the AGC was highly optimized, and arguably the most advanced computer of its time, as its computational power was only matched in the late 1970s by computers like the Apple II.

It is safe to say that the AGC had no intended market, and was explicitly designed to enhance control of the Apollo Command Module and Apollo Lunar Module. The AGC was not entirely internal to NASA, however, and was designed in MIT's Instrumentation Laboratory, and manufactured by Raytheon, a weapons and defense contractor.

\section{Analysis}

\subsection{ROM}

The Read-Only Memory (ROM) of the AGC is one of its defining features, and is surprisingly well-known even outside of groups that focus on ancient computing. This is primarily due to the distinctiveness of the chosen implementation: Core Rope Memory (CRM). 

In short, CRM involves the conditional wrapping of wire around magnetized ferrite cores in order to represent a 1 or a 0. If the wire passes through a core, that section of wire represents a 1, and if the wire passes around the core, that section of wire represents a 0. This memory scheme is incredibly space-efficient, as cores can be used by up to 24 wires, resulting in a dense mass of looping wire that spawned the name "rope". 

\fig{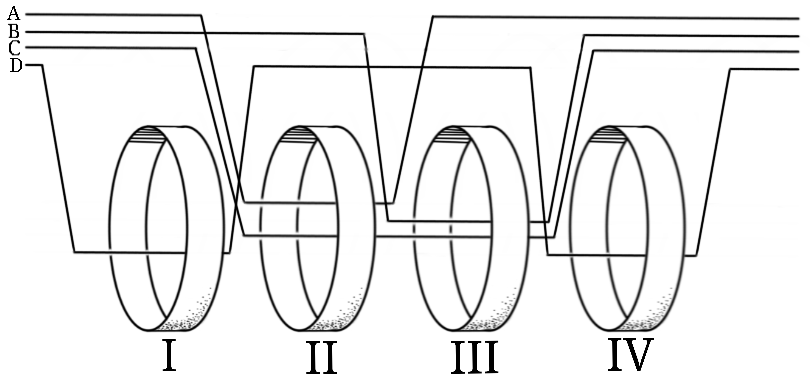}{Core Rope Memory}{1}

Here, fig. 1 shows an example of a simple CRM implementation that stores two bytes, and table 1 displays each line's value.

\begin{table}[h]
    \centering
    \begin{tabular}{||c c c c c||} 
        \hline
        & I & II & III & IV \\ [0.5ex] 
        \hline\hline
        A & 0 & 1 & 0 & 0 \\ 
        \hline
        B & 0 & 0 & 1 & 0 \\
        \hline
        C & 0 & 1 & 1 & 0 \\
        \hline
        D & 1 & 0 & 0 & 1 \\
        \hline
    \end{tabular}
    \textbf{\caption{Truth Table for Fig. 1}}
    \label{tab:1}
\end{table}

Although CRM was light, efficient, and reliable, it had a major drawback: it was handwoven. Because of this, single subroutines could take months to weave, and any error would be either irreparable or at the very least a huge setback.

The AGC used 72kb of ROM, however the 15-bit architecture did not support this many addresses. To absolve this issue, the memory was arranged into 36 memory banks that could be accessed one-at-a-time.

\clearpage

\subsection{RAM}

The Random Access Memory\footnote{Also referred to as "erasable memory"} (RAM) of the AGC functions under a similar concept as its ROM, using wires threaded through ferrite cores to represent bits. However, instead of threaded through/around representing a 1 or 0, each ferrite core is magnetized in one way to represent a 1, and can be magnetized in the opposite direction to represent a 0. This implementation is known as Magnetic Core Memory (MCM).

As all RAM is, MCM is designed to be read from and written to. Therefore, each core is threaded by a "Sense Line" that reads the polarization of the core's magnetic field, and a "Write Line" that can polarize the core either way. Cores are arranged in a grid, so one selection wire for each row and column are threaded through their respective cores. This threading scheme is illustrated in fig. 2.

\fig{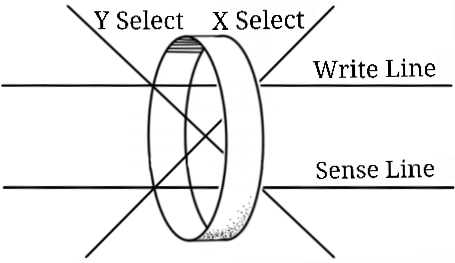}{Magnetic Core Memory}{2}

Although MCM was easier to manufacture (it was a standard erasable memory implementation at the time), its most significant limitation was that each bit of storage had a one-to-one mapping to a ferrite core. Cores were significantly heavier than wire and could not be reused for multiple bits, so the AGC's erasable memory held 18x less data than its ROM. Similar to its ROM, the AGC's RAM was arranged into 8 memory banks to resolve the 15-bit addressing issue.

The weight of the cores therefore had an indirect negative effect on the design of the AGC's software as well. Programmers were forced to optimize down to each individual bit in order to be able to store all necessary values in memory.

\subsection{Architecture}

The AGC used a "15 + 1"-bit \textbf{word} size, with the 16th bit being a \textbf{parity bit}. Any logical operation in the CPU explicitly operates on the 15 logical bits and ignores the parity bit. For error checking, the parity bit was set to a 1 or 0 such that the number of 1s in a word was odd, therefore if the number of 1s in a word was even, either data or instruction, then that word was corrupted.

In modern computers, registers are implemented as flip-flops, a very expensive form of memory that typically requires at least 20 transistors per bit. This is less of an issue in the era of transistors with sizes on the order of 2nm, but in the 1960s transistors were far larger. Therefore, only the 8 most important registers, the "central registers" were manufactured out of flip-flops, and the rest were memory mapped to MCM. Each of these central registers' intended purpose is described below.

\begin{itemize}
    \item \texttt{A ($0000_8$)} - The "accumulator", a general-purpose register used by almost all operations, contains data for most arithmetic and logical functions
    \item \texttt{L ($0001_8$)} - The "lower product register", \texttt{L} is a general-purpose register, however it is always used in conjunction with \texttt{A} when double precision is required
    \item \texttt{Q ($0002_8$)} - \texttt{Q} is used to store the return address of a function. As the AGC had no stack, nested functions were inherently prohibited unless \texttt{Q}'s value was stored in another register
    \item \texttt{EB ($0003_8$)} - \texttt{EB} stores a 3-bit field that determines which of the 8 RAM banks would be \textbf{memory-mapped} to the addresses $1400_8$-$1777_8$
    \item \texttt{FB ($0004_8$)} - \texttt{FB} serves the same function as \texttt{EB}, but uses a 5-bit field to map a ROM bank to the addresses $2000_8-3777_8$
    \item \texttt{Z ($0005_8$)} - \texttt{Z} stores the program counter, a value that contains the address of the next instruction to be executed once the current one completes
    \item \texttt{BB ($0006_8$)} - \texttt{BB} stores both a 3-bit field and a 5-bit field that allow a RAM bank and a ROM bank to be selected simultaneously
    \item \texttt{Unnamed ($0007_8$)} - This register is hardwired to the value 0, and is therefore utilized for reinitializing registers for computation. Note: this register is not implemented as a flip-flop but is still considered a central register
\end{itemize}

The AGC also used 40 more registers that were memory-mapped to RAM, but were not general-purpose and instead were used only for specific functions.

\subsection{Instruction Set}

The Block I AGC had an astonishing 12 instructions overall. These instructions covered basic arithmetic, conditional and unconditional jumps, a boolean AND, memory management, and even the ability to modify the upcoming instruction by adding a value in a register to its \textbf{opcode}.

Block II introduced 41 more instructions that provided significantly more functionality, including more complicated and optimized mathematical and logical functions, better interrupt handling, function returns, and more IO capabilities. Interestingly, the AGC's programmers also dedicated a significant amount of memory into designing an interpreter that could execute instructions not explicitly supported by the CPU.

\begin{center}
    \begin{minted}[xleftmargin=25pt,xrightmargin=10pt,linenos,frame=single]{nasm}
IGNITION
    ; INSURE ENGONFLG IS SET.
    CS      FLAGWRD5
    MASK    ENGONBIT
    ADS     FLAGWRD5
    ; TURN ON THE ENGINE.
    CS      PRIO30
    EXTEND
    RAND    DSALMOUT
    AD      BIT13
    EXTEND
    WRITE   DSALMOUT
    ; SET TEVENT FOR DOWNLINK
    EXTEND
    DCA     TIME2
    DXCH    TEVENT
    
    ; UPDATE TIG USING 
    ; TGO FROM S40.13
    EXTEND
    DCA     TGO
    DXCH    TIG
    EXTEND
    DCA     TIME2
    DAS     TIG
    \end{minted}
    \vspace{-1.5em}
    \textbf{\captionof{listing}{AGC Ignition Snippet}}
\end{center}

An example of the AGC's code is provided in Listing 1, specifically the beginning of the engine ignition subroutine.

Lines 3-5 load the 1's complement of flagword 5 into the accumulator, perform a logical AND on the accumulator with the "Engine On" bit, then add the result back to flagword 5 and save that sum back to its memory location. Here, flagword 5 and Engine On are individual bits that represent "flags" or "switches" that control simple \textbf{boolean} states. Essentially, this section of code confirms that the flag representing the state of the engine is on (this is also concisely summed up in the comment above).

The following section of code "TURN ON THE ENGINE." is harder to break down, as it uses the poorly-documented DSALMOUT IO register. However, the comment leads us to believe that in line 9, we "Read AND" mask the input register into the accumulator, add the contents of another undocumented BIT13, and write back out to DSALMOUT. It can be concluded that DSALMOUT is a memory-mapped IO register that controls power to the engine.

Following this section, we initialize some values for "DOWNLINK", or data transfer from the Lunar Command Module to Ground Control, and finally we clear and update some double-precision values (also undocumented).

\subsection{IO}

Aside from Core Rope Memory, the AGC's other defining feature is its \textbf{user interface}, DSKY (DiSplay and KeYboard).

Fig. 3 shows a diagram of DSKY's layout, obviously a long way from our displays and keyboards of today. 

The left section held light indicators for various states of the AGC, including whether or not it was at standby, whether it was in the middle of an uplink, status of the spacecraft, etc.

The right section held 7-segment displays that allowed the user to monitor values such as the spacecraft's velocity, required burn times, etc. More importantly, it displayed the current program being run and the data being operated on. For ease of understanding for the astronauts, programs were called "verbs", and data were called "nouns". The keyboard allowed astronauts to select a verb to execute and a noun to process, as well as input values to use for computations. 

A primitive \textbf{operating system} called "The Executive" would handle all of this input and use it to schedule programs, reset exceptions, and interact with the spacecraft.

DSKY, along with the AGC's other input devices such as an accelerometer or thermometer, were memory-mapped to registers in RAM for use in computation.

\fig{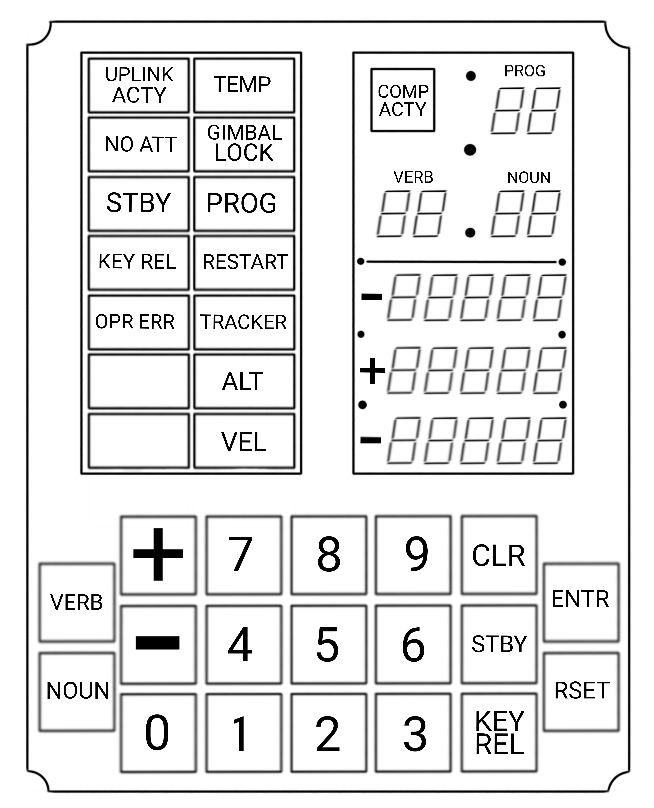}{DSKY Interface}{3}

\clearpage

\section{Reviews}

\subsection{}

\textbf{"In Defense of the Apollo Program's Guidance Computer" - John Loeffler}

The phrase "your calculator is more powerful than the computer that piloted the Apollo spacecraft" is somewhat common, and many people who hear it would agree with the statement. 

Loeffler partially disagrees; while the statement is \textit{technically} true, it completely ignores the immense amount of engineering that went into the AGC. Most of its technology was completely novel, and many of its software perks were "ahead of [their] time by decades". To claim the AGC was not a powerful computer abuses the definition of the word.

\subsection{}

\textbf{"Rebooting a 50 Year Old Computer - Making The Apollo Guidance Computer Work Again" - Scott Manley}

This repair video starts off with an excellent review of the AGC's significance. Manley reiterates common points about the ingenuity of the technology within the AGC, but brings special attention to its impact on technology as a whole. 

During the development of Block I, silicon transistors and embedded circuit technology became feasible enough to use in computers. Block I was able to transition from racks of large-scale transistors hand-soldered together to a smaller implementation. As silicon technology became essential to keeping the AGC small, there is evidence to believe that the demand for silicon transistors generated by NASA during the AGC's design period was so great that it kickstarted general-purpose computing and personal computing in the early 1970s. Therefore, our modern computers and phones have direct ties to the AGC.

\subsection{}

\textbf{"The Real Story Behind the Apollo 11 Computer Error" - Wall Street Journal}

When descending to the surface of the Moon for the first time, the AGC flashed an error code that perplexed astronauts and Ground Control. WSJ interviews Don Eyles, the MIT Instrumentation Laboratory employee who wrote the code to control the Lunar descent.

Although the AGC was erroring without a known cause, Ground Control soon discovered that it was still executing critical guidance and maneuvering programs, and the Lunar Module was able to successfully land without a hitch. WSJ claims that this feat attests to the robustness of the AGC's software, that it was able to survive an unknown issue during one of the most critical points of the mission.

It was later discovered that the issue was not even within the AGC, but that a radar dish had been activated, filling RAM with unneeded data. Despite this, the AGC would reboot after dumping its memory, and upon booting again would resume from where it left off. This is a feature that is uncommon even in much of today's software, so the fact that it was not only present but functional in a computer from the 1960s is an incredible feat of engineering.

\clearpage

\section{Technical Terms}

\begin{itemize}
    \item{
        \textbf{AGC}
        
        Apollo Guidance Computer
    }
    \item{
        \textbf{Boolean}
        
        A value that is either True or False, represented as a 1 or 0 in a computer
    }
    \item{
        \textbf{FLOPS}
        
        "Float Operations Per Second" - a unit of measurement for measuring computer speed based on the number of single-precision multiplications and additions the computer can compute in one second
    }
    \item{
        \textbf{Memory-Mapped}
        
        Memory-Mapping is a technique that maps some input to a given memory address. For example, memory-mapped keyboard input might place the ASCII value of the last key typed into memory address \texttt{0xABCD}
    }
    \item{
        \textbf{Opcode}
        
        An opcode is a sequence of bits unique to an instruction that tell the CPU how to execute the instruction
    }
    \item{
        \textbf{Operating System}
        
        An Operating System, or OS, is a central piece of software in a computer that controls everything the user interacts with, including but not limited to program execution, keyboard input, display, etc.
    }
    \item{
        \textbf{Parity Bit}
        
        A parity bit is a single bit of data at the end of each word of memory used for error detection.
    }
    
    \newpage
    
    \item{
        \textbf{RAM}
        
        "Random Access Memory" - Computer memory that can be read from and written to. In the AGC, this was implemented with Magnetic Core Memory and is called "erasable memory" in documentation
    }
    \item{
        \textbf{ROM}
        
        "Read-Only Memory" - Computer memory that can be read from, but not written to. In the AGC, this was implemented with Core Rope Memory
    }
    \item{
        \textbf{User Interface}
        
        The User Interface, or UI, is what the user interacts with when operating a computer. This includes input devices such as a mouse and keyboard, and output devices such as a display.
    }
    \item{
        \textbf{Word}
        
        A word is a unit of measurement defined by the minimum amount of bits the CPU may operate on. In the AGC, word size is 16, 15 bits for data and 1 parity bit.
    }
\end{itemize}

\onecolumn

\section{References}

\begin{itemize}
    \item O'Brien, F. (2010). The apollo guidance computer: Architecture and operation. Springer. 
    
    \item Hall, E. C. (1963, May). General Design Characteristics of the Apollo Guidance Computer. Boston; MIT Instrumentation Laboratory. \\
    \url{http://klabs.org/history/history_docs/mit_docs/1009.pdf}
    
    \item Blair-Smith, H. (1966). AGC4 Memo \# 9 - Block II Instructions. Boston; MIT Instrumentation Laboratory. \\
    \url{http://authors.library.caltech.edu/5456/1/hrst.mit.edu/hrs/apollo/public/archive/1689.pdf}
    
    \item Burkey, R. (n.d.). Programmer's Manual - Block 2 AGC Assembly Language. Virtual AGC assembly-Language manual. Retrieved December 2, 2021.\\
    \url{https://www.ibiblio.org/apollo/assembly_language_manual.html}. 
\end{itemize}

\end{document}